\documentclass[sn-mathphys,Numbered]{sn-jnl}

\usepackage{graphicx}%
\usepackage{multirow}%
\usepackage{amsmath,amssymb,amsfonts}%
\usepackage{amsthm}%
\usepackage{mathrsfs}%
\usepackage[title]{appendix}%
\usepackage{xcolor}%
\usepackage{textcomp}%
\usepackage{manyfoot}%
\usepackage{booktabs}%
\usepackage{algorithm}%
\usepackage{algorithmicx}%
\usepackage{algpseudocode}%
\usepackage{listings}%

\usepackage{gensymb,adjustbox,siunitx,threeparttable,setspace,xr,bm}
\begin{document}

\title[Article Title]{Parsing altered brain connectivity in neurodevelopmental disorders by integrating graph-based normative modeling and deep generative networks}


\author[1,2]{\fnm{Rui Sherry} \sur{Shen}}

\author[3]{\fnm{Yusuf} \sur{Osmanlıoğlu}}

\author[2]{\fnm{Drew} \sur{Parker}}

\author[2]{\fnm{Darien} \sur{Aunapu}}

\author[4,5,6]{\fnm{Benjamin E.} \sur{Yerys}}

\author[4,5]{\fnm{Birkan} \sur{Tunç}}

\author[2]{\fnm{Ragini} \sur{Verma}}

\affil[1]{\orgdiv{Department of Bioengineering}, \orgname{University of Pennsylvania}}

\affil[2]{\orgdiv{DiCIPHR lab, Department of Radiology, Perelman School of Medicine}, \orgname{University of Pennsylvania}}

\affil[3]{\orgdiv{Department of Computer Science}, \orgname{Drexel University}}

\affil[4]{\orgdiv{Center for Autism Research}, \orgname{Children’s Hospital of Philadelphia}}

\affil[5]{\orgdiv{Department of Child and Adolescent Psychiatry and Behavioral Science}, \orgname{Children’s Hospital of Philadelphia}}

\affil[5]{\orgdiv{Advancing Transition and Learning for Adult Success Center}, \orgname{Children’s Hospital of Philadelphia}}


\abstract{
Divergent brain connectivity is thought to underlie the behavioral and cognitive symptoms observed in many neurodevelopmental disorders. Quantifying divergence from neurotypical connectivity patterns offers a promising pathway to inform diagnosis and therapeutic interventions. While advanced neuroimaging techniques, such as diffusion MRI (dMRI), have facilitated the mapping of brain's structural connectome, the challenge lies in accurately modeling developmental trajectories within these complex networked structures to create robust neurodivergence markers. In this work, we present the Brain Representation via Individualized Deep Generative Embedding (BRIDGE) framework, which integrates normative modeling with a bio-inspired deep generative model to create a reference trajectory of connectivity transformation as part of neurotypical development. This will enable the assessment of neurodivergence by comparing individuals to the established neurotypical trajectory. BRIDGE provides a global neurodivergence score based on the difference between connectivity-based brain age and chronological age, along with region-wise neurodivergence maps that highlight localized connectivity differences. Application of BRIDGE to a large cohort of children with autism spectrum disorder demonstrates that the global neurodivergence score correlates with clinical assessments in autism, and the regional map offers insights into the heterogeneity at the individual level in neurodevelopmental disorders. Together, the neurodivergence score and map form powerful tools for quantifying developmental divergence in connectivity patterns, advancing the development of imaging markers for personalized diagnosis and intervention in various clinical contexts.
}




\maketitle

\section{Introduction}\label{sec1}

The human brain can be viewed as a complex networked structure that undergoes radical transformations to support increasingly sophisticated cognitive, social, and motor skills, all while maintaining relatively low metabolic costs. Many neurodevelopmental disorders, including autism spectrum disorder (ASD), attention deficit hyperactivity disorder (ADHD), and schizophrenia, have long been linked to divergent patterns of connectivity within the brain \cite{belmonte2004autism,uddin2013reconceptualizing,konrad2010adhd,rapoport2005neurodevelopmental}, which manifest as a range of cognitive and behavioral symptoms. Quantification of individual-level neurodivergence in brain connectivity is expected to transform clinical decision-making and parse heterogeneity in neurodevelopmental disorders, thereby making targeted therapeutics a reality.

Advances in neuroimaging, especially diffusion MRI (dMRI) \cite{zhang2022quantitative}, have enabled an in vivo reconstruction of the brain's structural connectivity, known as the structural connectome, with nodes representing brain regions and edges denoting white matter connections. This progress has led to a surge of studies investigating connectivity-based changes in neurodevelopmental disorders \cite{rudie2013altered,lewis2014network,wheeler2014review}, predominantly utilizing a case-control paradigm that focuses on group average differences between the neurodivergent group and matched neurotypicals. While these studies have provided considerable insights into connectivity patterns associated with symptoms of neurodevelopmental disorders and have strengthened the hypothesis that structural connectomes are affected, their design inherently overlooks individual differences and variations within groups. Consequently, the case-control approach falls short, as most neurodevelopmental disorders exhibit substantial heterogeneity in various developmental stages \cite{khodosevich2023neurodevelopmental}, diverse clinical presentations \cite{fountain2012six}, and underlying mechanisms \cite{jeste2014disentangling}. As a result, group-based approaches often yield inconclusive findings, such as reports of both hypo- and hyper-connectivity in autism \cite{vissers2012brain,ha2015characteristics}.  In our work, we focus on creating individual-level brain connectivity markers of neurodivergence to support precision medicine.

A major roadblock in developing imaging markers for neurodevelopmental disorders is the need to robustly quantify their high levels of heterogeneity. Normative modeling approaches have emerged as a solution, enabling the characterization of variability in neurotypical development and providing individual-level statistical inference to quantify the degree of neurodivergence \cite{marquand2019conceptualizing,marquand2016understanding,tuncc2019deviation}. Much like growth charts in pediatric medicine, which compare a child’s body measurements to typical age-specific variability, a normative model of brain development characterizes neurodivergent individuals by comparing their brain features to established reference ranges of neurotypical developmental trajectories \cite{rutherford2022charting}. Researchers have applied normative modeling to probe developmental heterogeneity based on cortical thickness \cite{bethlehem2020normative,wolfers2018mapping}, brain volume \cite{wolfers2020individual}, white matter integrity measures \cite{tuncc2019deviation} and other derived brain measures \cite{floris2021atypical,lv2021individual}.

However, traditional normative models cannot tackle the complexity of brain connectivity, which requires to extract network-based information encoded in the connectomes. Employing normative modeling directly on connectivity values at the edges of the structural connectomes may overlook the interdependent topological properties of the network. When viewed as a graph, graph-theoretical measures \cite{bullmore2009complex} of the structural connectome can describe certain topological features, but those pre-defined measures fail to provide an adequate characterization and are generally insensitive to complex development-related transformations \cite{betzel2016generative}. Generative network models offer a promising alternative to encode connectivity information through parameterized wiring rules that mimic biological factors involved in brain maturation \cite{kaiser2004modelling,bullmore2012economy,betzel2016generative,betzel2017generative}. Additionally, recent deep generative models, such as graph variational autoencoders (VAE) \cite{kingma2013auto,kipf2016variational}, provide powerful data-driven approaches for extracting valuable features, enabling efficient characterization of developmental transformations from high-dimensional structural connectome data.

We present a novel {\bf B}rain {\bf R}epresentation via {\bf I}ndividualized {\bf D}eep {\bf G}enerative {\bf E}mbedding (BRIDGE) framework to address the absence of normative models on connectomes for the investigation of neurodevelopmental disorders. BRIDGE uniquely integrates deep generative models and normative modeling for the first time. It incorporates biologically relevant wiring constraints to guide the generation of structural connectomes and uses graph VAE to infer latent brain configurations, effectively characterizing connectivity patterns during development. We demonstrate the superiority of BRIDGE in modeling the neurotypical development and highlight its utility in detecting neurodivergence in children with autism. Our approach provides neurodivergence maps that pinpoint regional connectivity differences at the individual level, along with global neurodivergence scores that correlate with symptom measures. Our findings underscore the potential of BRIDGE to reveal the intricate dynamics of brain connectivity development, paving the way for imaging markers that enable more personalized diagnosis and intervention.

\section{Results}\label{sec2}
\begin{figure*}[h]
    \centering
	\includegraphics[width=1\linewidth]{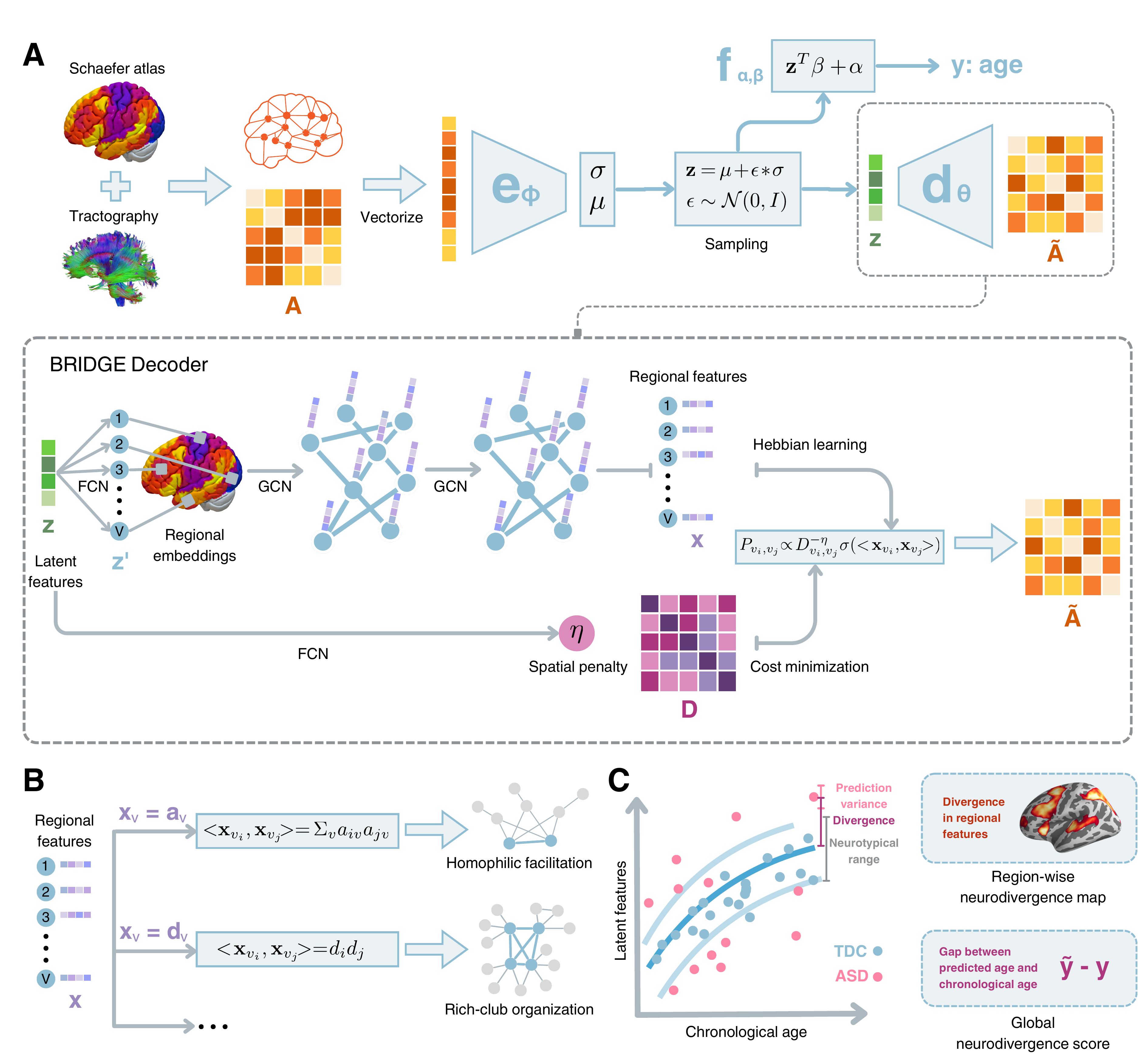}
	\caption{Overview of the BRIDGE framework. A) Architecture of the BRIDGE model for learning neurotypical developmental trajectories of structural connectivity using connectomes. Graph-based VAE was used to encode the structural connectivity information in connectome via lower-dimensional latent features. The BRIDGE decoder mapped these latent features to each brain region and inferred the regional features to generate structural connectomes. This generative process was guided by two biological wiring constraints, Hebbian learning and cost minimization. B) Illustration showing how two properties of brain connectivity, homophilic facilitation and rich-club organization, could be replicated by the BRIDGE model via specific regional features. C) Derivation of the region-wise neurodivergence map and the global neurodivergence score.}
	\label{F1}
\end{figure*}

As described in the Methods, experiments were designed to 1) establish that the BRIDGE framework, as illustrated in Fig. \ref{F1}, provides a superior normative model for structural connectomes compared to other ways of encoding developmental changes in connectivity (Section \ref{sec2:a} and \ref{sec2:b}); and 2) validate that the region-wise neurodivergence map (Section \ref{sec2:c}) and the global neurodivergence score (Section \ref{sec2:d}), derived with BRIDGE as the normative reference, offer clinically meaningful measures that are more robust than those from other models. Table \ref{table0} presents the demographics of the two datasets used in this study: the Philadelphia Neurodevelopmental Cohort (PNC) \cite{satterthwaite2014neuroimaging} and the Center for Autism Research (CAR) cohort \cite{ghanbari2014identifying}.
\begin{table*}[h]
\caption{Demographics of PNC and CAR datasets}
\begin{adjustbox}{width=\textwidth,center}
    \begin{tabular}{c|c|c|c}
        \multicolumn{1}{c|}{} & \multicolumn{1}{c|}{PNC} & \multicolumn{2}{c}{CAR}\\
        \toprule
        Diagnoses & NT=968 & NT=196 & ASD=229\\
        \hline
        Sex & M=425, F=54 & M=158, F=38 & M=190, F=39\\
        \hline
        Age & $15.32 \pm 3.45$ (8-22) & $12.37 \pm 2.96$ (6-19) & $12.46 \pm 2.72$ (6-19) \\
        \hline
        IQ & --- & $113.81 \pm 16.02$ (75-155) & $100.92 \pm 21.99$ (47-165) \\
        \hline
        ADOS & --- & $1.60 \pm 1.38$ (1-7) & $7.13 \pm 1.99$ (1-10) \\
        \bottomrule
    \end{tabular}
    \label{table0}
\end{adjustbox}
\footnotesize{Abbreviations: NT: neurotypical, ASD: autism spectrum disorder, M: male, F: female, IQ: DAS-II intelligence quotient, ADOS: Autism Diagnostic Observation Schedule calibrated severity score.\\Mean, variance, and ranges are reported for age (in years)/IQ/ADOS at scan.}\\
\end{table*}

\subsection{BRIDGE captured topological properties in structural connectomes} \label{sec2:a}

\begin{figure*}[h]
    \centering
	\includegraphics[width=0.9\linewidth]{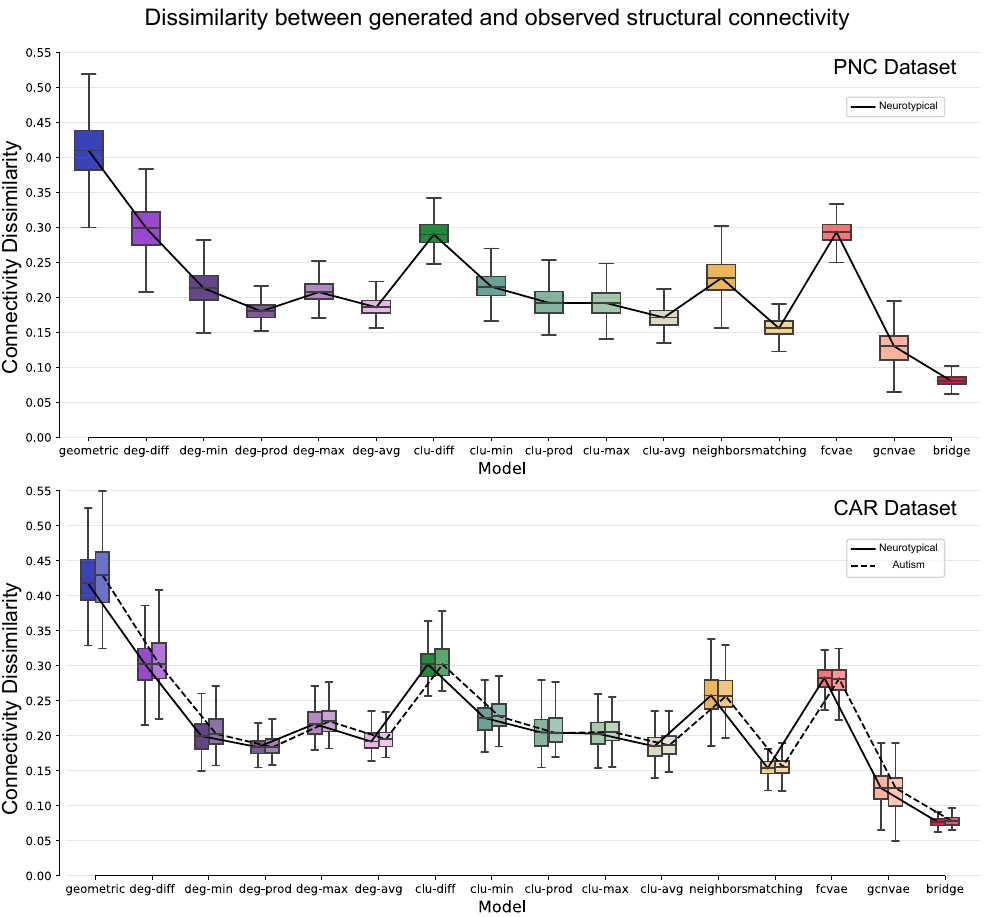}
	\caption{Performance of the BRIDGE model were compared with two other deep generative models (FC-VAE, GCN-VAE) and 13 classic generative network models based on the dissimilarity between generated and actual structural connectomes. For the CAR dataset, dissimilarity was computed separately for the neurotypical group and the autism group. The BRIDGE model achieved the lowest dissimilarity, positioning it as the optimal choice for both datasets irrespective of diagnostic status.}
	\label{R1}
\end{figure*}

The BRIDGE framework extracted lower-dimensional latent features from the structural connectomes and modeled the generative process of the structural connectomes across different ages. We anticipated that connectomes generated by the BRIDGE model would exhibit graph topological properties similar to those of actual structural connectomes. Results of the overall dissimilarity in graph topological properties are shown in Fig. \ref{R1}, where the BRIDGE model was compared with other deep learning-based and classic generative models as described in the Methods. The BRIDGE model achieved the best performance in capturing connectivity information, with the lowest dissimilarity between generated and actual structural connectomes across both datasets, followed by GCN-VAE and the matching index model. Incorporating biological wiring constrains into the BRIDGE model significantly improved the generation of structural connectomes, compared to GCN-VAE with paired t-tests for both the PNC (Cohen's $d=2.46, p<0.001$) and CAR (neurotypical: Cohen's $d=2.69, p<0.001$, autism: Cohen's $d=1.82, p<0.001$) datasets. When the same analysis was repeated after threshlding structural connectomes at different levels of sparsity ($\rho=10\%, 20\%, 40\%$), the BRIDGE model maintained the highest performance in capturing underlying connectivity information. No significant group differences in dissimilarity metrics were observed between autistic and neurotypical participants, suggesting that the BRIDGE model is equally effective at fitting structural connectomes regardless of diagnostic status.

\subsection{BRIDGE encoded age-related connectivity patterns in normative modeling} \label{sec2:b}

\begin{figure*}[h]
    \centering
	\includegraphics[width=0.9\linewidth]{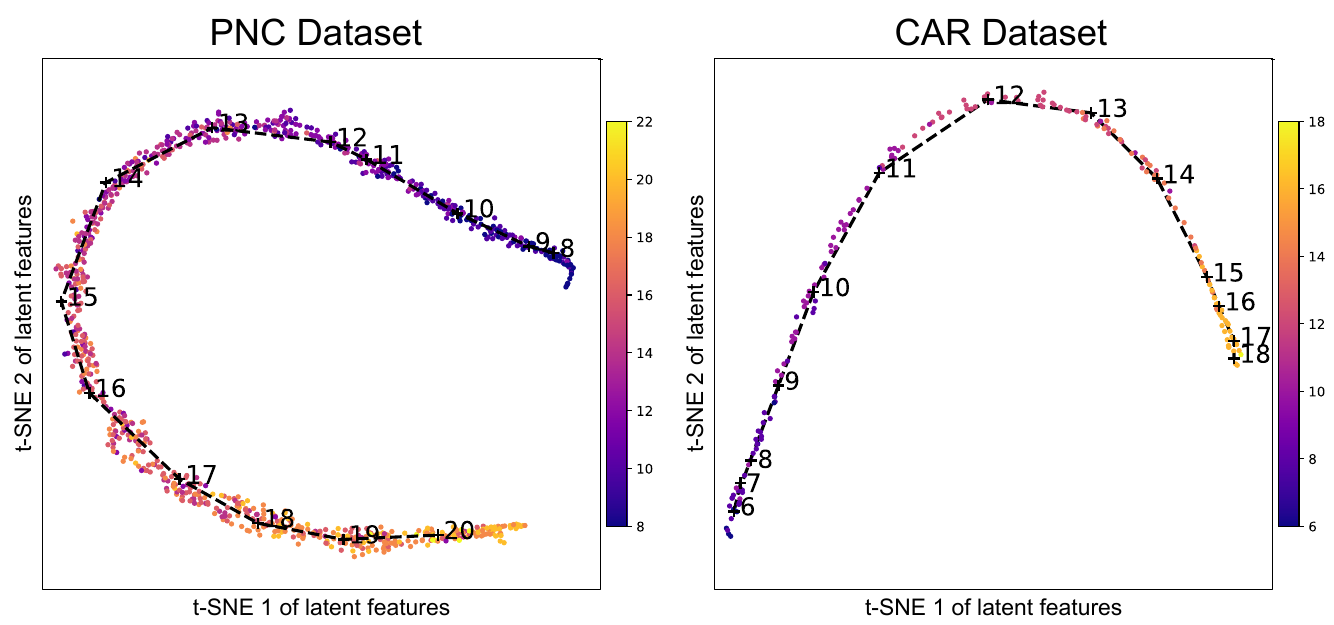}
	\caption{Visualization of latent space using t-SNE. Latent features derived from BRIDGE showed a notable age-related trend for neurotypical participants in both the PNC (left) and CAR (right) cohorts. Feature dimensions were reduced to 2D using t-SNE for visualization. Each dot was colored according to the participant's chronological age. The average trajectories of neurotypical participants across different ages were represented using as the black dashed lines.}
	\label{R2}
\end{figure*}

\begin{table*} [h]
\caption{Comparison of brain age prediction on neurotypicals}
\begin{adjustbox}{width=\textwidth,center}
    \begin{tabular}{c|c|c|c|c|c|c|c|c|c}
        \multicolumn{1}{c|}{} & \multicolumn{3}{c|}{PNC} & \multicolumn{3}{c|}{CAR (w/o pretrain)} & \multicolumn{3}{c}{CAR (w/ pretrain)}\\
        \toprule
        \multicolumn{1}{c|}{Models} & \multicolumn{1}{c|}{MAE} & \multicolumn{1}{c|}{RMSE} & \multicolumn{1}{c|}{R} & \multicolumn{1}{c|}{MAE} & \multicolumn{1}{c|}{RMSE} & \multicolumn{1}{c|}{R} & \multicolumn{1}{c|}{MAE} & \multicolumn{1}{c|}{RMSE} & \multicolumn{1}{c}{R}\\
        \hline
        $\bar{y}_{age}$ (Baseline)  & 2.8969 & 3.4384& --- & 2.4701 & 2.9264 & --- & --- & --- & --- \\
        \hline
        LR-PCA & 2.1115 & 2.6122 & 0.6535 & 1.5572 & 1.9778 & 0.7328 & --- & --- & --- \\
        \hline
        SVR-PCA & 2.0910 & 2.5995 & 0.6644 & 1.6284 & 2.0409 & 0.7269 & --- & --- & --- \\
        \hline
        MLP & 2.0087 & 2.4890 & 0.7078 & 1.5614 & 1.9194 & 0.7759 & 1.3878 & 1.7794 & 0.8098 \\
        \hline
        FC-VAE & 1.9514 & 2.4237 & 0.7105 & 1.5750 & 1.9702 & 0.7863 & 1.3869 & 1.7602 & 0.8160 \\
        \hline
        GCN-VAE & 1.9685 & 2.4110 & 0.7151 & 1.5306 & 1.8842 & 0.7990 & 1.4154 & 1.7472 & 0.8293 \\
        \hline
        BRIDGE & \textbf{1.8751} & \textbf{2.3230} & \textbf{0.7375} & \textbf{1.4890} & \textbf{1.8621} & \textbf{0.8110} & \textbf{1.3438} & \textbf{1.6841} & \textbf{0.8312} \\
        \bottomrule
    \end{tabular}
    \label{table1}
\end{adjustbox}
\footnotesize{$^1$Two settings were considered when testing on the CAR cohort: random weight initialization for deep neural networks (w/o pretrain) and initialization with weights pre-trained from the PNC cohort (w/ pretrain)\\$^2$MAE: mean absolute error, RMSE: root mean square error, R: Pearson's correlation\\$^3$The presented results were based on 5-fold cross-validation}\\
\end{table*}

\begin{figure*}[h]
    \centering
	\includegraphics[width=1\linewidth]{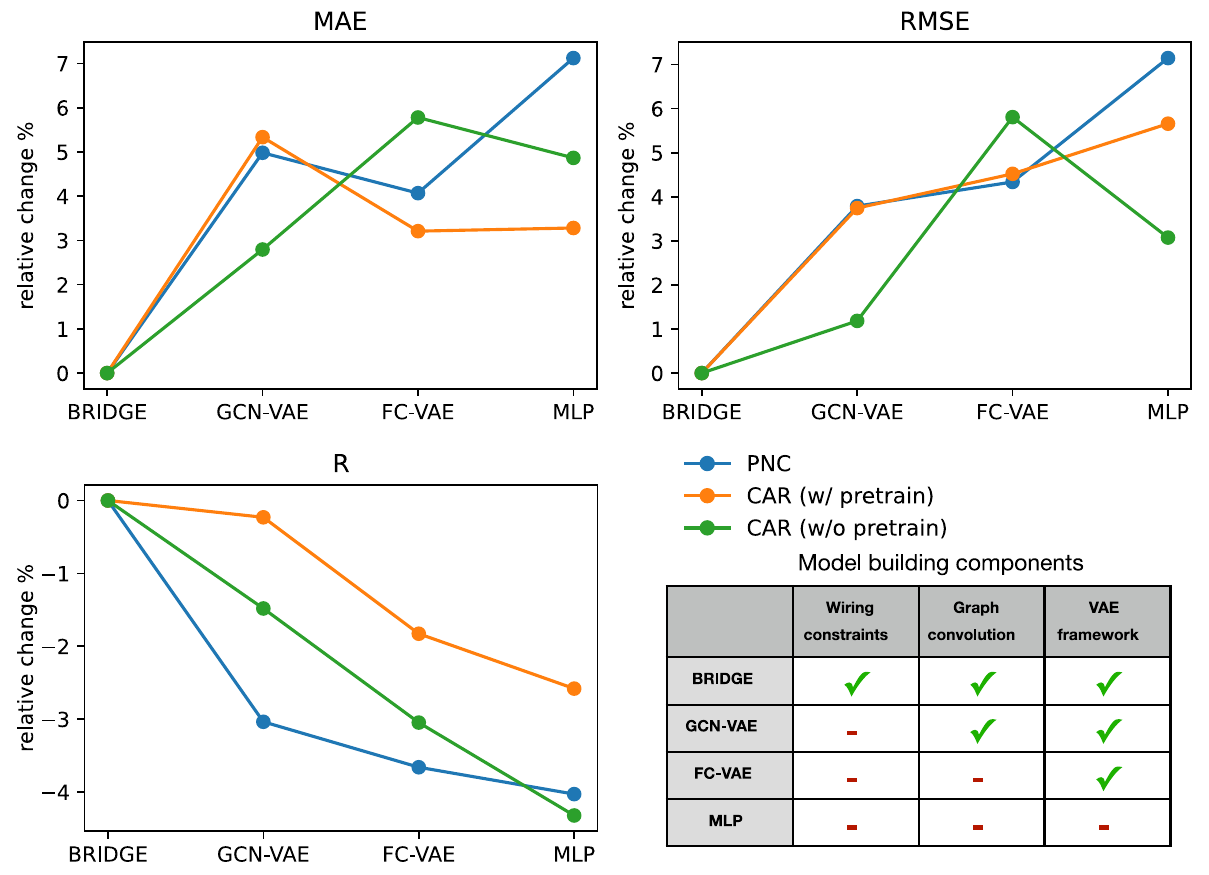}
	\caption{Ablation study by removing each building component in BRIDGE. The constituent elements of BRIDGE and three competing models were illustrated in the table. Comparative assessment of percentage changes upon removal of each building components was evaluated through MAE, RMSE, and R metrics.}
	\label{R3}
\end{figure*}

With the goal of demonstrating that the latent features of the BRIDGE model encode age-related changes in connectivity more effectively than other normative model, we utilized BRIDGE and other models to create connectome-based predictors of age, referred to as ``brain age''.

In Fig.\ref{R2}, we visualized the BRIDGE latent features in two dimensions using t-SNE \cite{van2008visualizing}, with each dot's color representing the chronological age of each neurotypical participant, and the black dashed lines indicating the averaged latent features across the neurotypicals at each age. The encoded latent features displayed a clear age-related trend in both datasets, suggesting that age-related characteristics were effectively captured. 

Table \ref{table1} compares the performance of BRIDGE in predicting brain age from structural connectomes with various brain age predictors, including linear regression with principal component analysis (LR-PCA), support vector regression with PCA (SVR-PCA), multilayer perceptrons (MLP), standard VAE models with fully connected networks (FC-VAE) and graph convolutional networks (GCN-VAE), as well as a baseline model using the mean chronological age as the predicted brain age for all individuals. We demonstrated that BRIDGE reduced prediction errors by 32\% - 40\% from the baseline model, by 11\% - 13\% from classic machine learning models (LR-PCA and SVR-PCA), and by 3-9\% from deep learning models (MLP, FC-VAE, and GCN-VAE). Additionally, we found that pre-training the model using the larger PNC cohort remarkably improved brain age prediction in the CAR cohort. Collectively, our results showed that BRIDGE consistently outperformed all other models.

Specifically, our comparison of BRIDGE with MLP, FC-VAE, and GCN-VAE revealed that integrating three key components (the VAE architecture, graph convolutional layers, and biological wiring constraints) into the BRIDGE model improved brain age prediction. The percentage of changes after removing each building components is visualized in Fig.\ref{R3}. Notably, the addition of wiring constraints (BRIDGE versus GCN-VAE) markedly reduced the prediction errors and improved correlation with chronological age. Similarly, incorporating the graph convolutional layers (GCN-VAE versus FC-VAE) and the deep generative framework (FC-VAE versus MLP) boosted the predictive ability in most scenarios.

In contrast, classic generative models based on pre-defined graph features \cite{betzel2016generative,betzel2017generative} failed to effectively encode developmental changes in structural connectomes. Table S\ref{sup-table1} shows the Pearson correlation between chronological age and the classic generative model parameters for neurotypical participants. Among the 13 models, features derived from degree-based models (except for deg-diff) and certain clustering coefficient-based models exhibited a weak correlation with age in the PNC cohort ($\rho<0.2, p<0.05$, FWER corrected). In the CAR cohort, a significant correlation with chronological age was observed only for the $\eta$ feature (which controls the penalty on connection length) from the deg-avg model ($\rho=-0.2437, p<0.05$, FWER corrected). None of the brain age predictors based on features from classic generative model showed a noticeable improvement in predicting brain age compared to the baseline, as shown in Table S\ref{sup-table2}, indicating that these classic models are inadequate for characterizing age-related connectivity trajectories.

Moreover, we compared the performance of BRIDGE and other models in predicting brain age among autistic participants, as shown in Table S\ref{sup-table3}. We observed that the divergent connectivity patterns in autism led to reduced Pearson correlation and increased prediction errors compared to neurotypical participants. Consistent with the findings in the neurotypicals, the BRIDGE model outperformed all other brain age predictors across all metrics.

\subsection{Regional neurodivergence maps revealed individualized differences for neurodevelopmental disorders} \label{sec2:c}

\begin{figure*}[h!]
    \centering
	\includegraphics[width=0.6\linewidth]{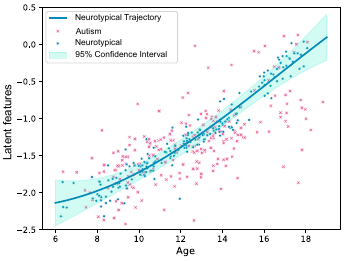}
	\caption{Illustration of neurodivergence in one dimension of latent features captured by the BRIDGE model. Neurotypical participants are represented by blue dots, while autistic children are marked with red "X"s. The neurotypical trajectory across different chronological ages is shown as a blue line, enveloped by a shaded area representing the 95\% confidence interval, reflecting the variability among neurotypicals. Autistic children display divergent developmental patterns in their latent features compared to neurotypicals.}
	\label{R4}
\end{figure*}

\begin{figure*}[h!]
    \centering
	\includegraphics[width=0.85\linewidth]{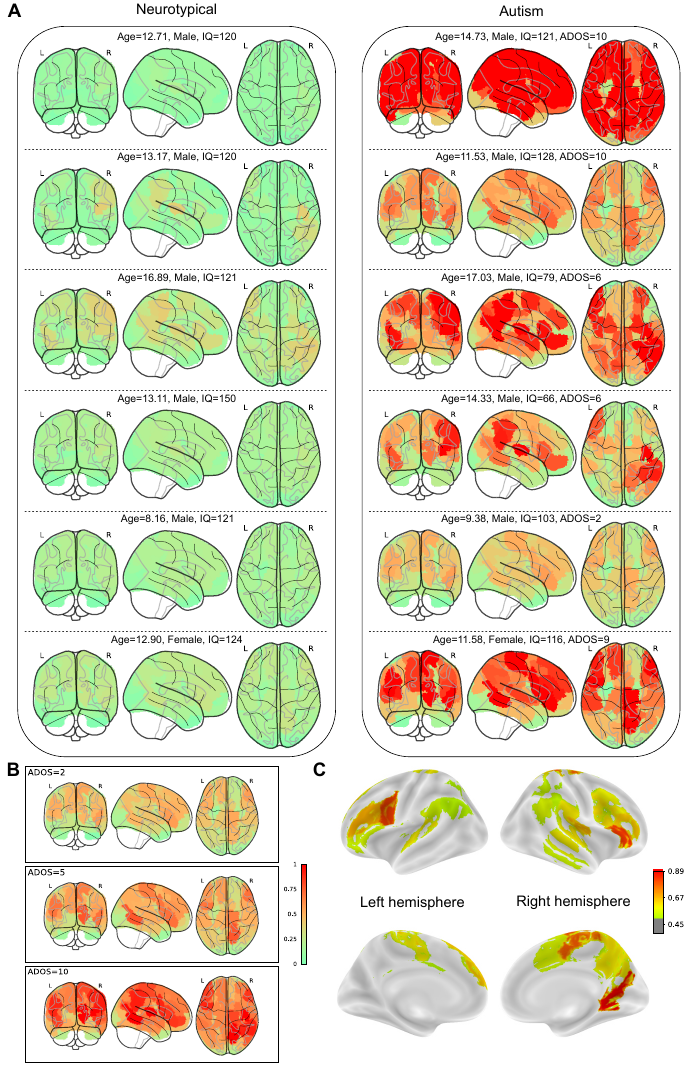}
	\caption{Individualized regional neurodivergence maps revealed heterogeneity in structural connectome phenotypes for autism. A) Regional neurodivergence maps for 12 participants, along with age, sex, IQ, and autistic traits. B) Averaged regional neurodivergence maps for participants with an ADOS score of 2, 4, and 10. C) The identification of brain regions commonly associated with autism.}
	\label{R5}
\end{figure*}
We mapped the latent features derived from the structural connectomes for each brain region to examine region-specific neurodivergence in each participant. Fig. \ref{R4} illustrates one dimension of the latent features derived from BRIDGE, with blue dots denoting neurotypical participants and red `X' markers indicating autistic participants. The blue line shows the neurotypical trajectory across the age range, enveloped by a shaded area representing the 95\% confidence interval. Neurotypical participants predominantly fell within this established range, whereas most autistic children lay outside the 95\% confidence interval.

We derived region-wise divergences across the entire brain as detailed in the Methods, resulting in individualized regional neurodivergence maps, as shown in Fig. \ref{R5}A for 12 example participants. As expected, autistic children showed greater regional neurodivergence from neurotypical trajectories. The patterns of regional neurodivergence varied among autistic participants, even for those with similar IQ and ADOS scores, highlighting substantial heterogeneity in their structural connectome phenotypes. Fig. \ref{R5}B presents the average region-wise neurodivergence maps for participants with Autism Diagnostic Observation Schedule (ADOS) calibrated severity scores of 2, 4, and 10, respectively. Individuals with higher ADOS scores exhibited greater regional neurodivergence and a broader involvement of brain regions compared to those with lower scores. We averaged the region-wise neurodivergence maps across all autistic participants to identify regions that are most frequently divergent in this autism cohort, as shown in Fig. \ref{R5}C. The top 10 regions with the highest neurodivergence were the visual network (right medial visual area), default mode network (bilateral prefrontal area), dorsal attention network (bilateral frontal eye fields), somatomotor network (right pericentral area), and executive control network (bilateral lateral prefrontal area).

\subsection{Clinical importance of the global neurodivergence scores} \label{sec2:d}

\begin{figure*}[h!]
    \centering
	\includegraphics[width=0.9\linewidth]{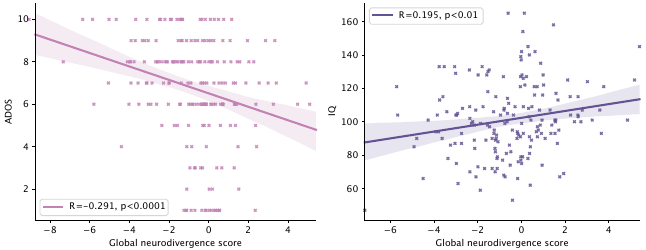}
	\caption{The global neurodivergence scores captured by the BRIDGE model significantly correlated with ADOS (left) and IQ (right) scores.}
	\label{R6}
\end{figure*}

We investigated the significance of global neurodivergence scores, defined as the difference between predicted brain age and chronological age, in capturing clinically meaningful aspects of neurodevelopmental disorders. Global neurodivergence scores derived from BRIDGE were significantly correlated with ADOS (Spearman $R=-0.291, p<0.0001$, 95\% CI=[-0.414, -0.158]) and IQ scores (Pearson $R=0.195, p<0.01$, 95\% CI=[0.049, 0.333]), as shown in Fig.\ref{R6}. The correlation between global neurodivergence scores and ADOS remained significant after controlling for IQ (Spearman $R=-0.221, p<0.01$, 95\% CI=[-0.352, -0.083]), indicating that these neurodivergence scores captured both cognitive and autistic aspects.

We found that GCN-VAE and FC-VAE models produced neurodivergence scores with weaker correlations with ADOS (GCN-VAE: $R=-0.209, p<0.005$, 95\% CI=[-0.340, -0.070]; FC-VAE: $R=-0.211, p<0.005$, 95\% CI=[-0.342, -0.072]) and IQ (GCN-VAE: Pearson $R=0.142, p=0.059$, 95\% CI=[-0.005, 0.284]; FC-VAE: Pearson $R=0.184, p<0.05$, 95\% CI=[0.038, 0.323]). No significant relationships were identified using the MLP, SVR-PCA, or LR-PCA models.

\section{Discussion} \label{sec3}
The results presented a comprehensive view of our BRIDGE framework that uniquely combines normative modeling with a biologically constrained graph VAE to encode neurotypical developmental patterns of structural connectivity for the first time. Through BRIDGE, we introduced regional and global connectivity markers to assess neurodivergence at the individual level.

Normative modeling of structural connectomes poses challenges due to the complexity of encoding high-dimensional connectome data while preserving their underlying topological properties in the model. BRIDGE extracted low-dimensional latent features through learning the generation of structural connectomes across various ages with a graph VAE architecture, by uniquely incorporating biologically relevant wiring constraints. Standard deep generative models, like FC-VAE, struggle to capture graph-based information and offer limited biological interpretability due to their "black box" nature \cite{markus2021role, reyes2020interpretability}. Although the latent features extracted by FC-VAE effectively predicted brain age, they fell short in characterizing desired topological properties of structural connectomes. A side-by-side comparison with GCN-VAE emphasized the critical role of graph convolutional layers in handling graph-structured data, particularly in scenarios with limited sample sizes \cite{zhou2020graph}. Our BRIDGE framework further refined the generative process by integrating two wiring principles: Hebbian learning and cost minimization, both of which were well-established in various biological processes \cite{niven2008energy,buzsaki2004interneuron,caporale2008spike,keysers2014hebbian}. Incorporating wiring constraints remarkably boosted the model's fit to structural connectomes and brain age prediction compared to other VAE-based approaches. These wiring constraints guided the model in distilling insightful latent features that aligned with biological and mechanistic processes, thereby enabling effective normative modeling that captured age-related neurotypical variability in structural connectomes. 

We also compared BRIDGE with classic two-parameter generative models using pre-defined graph measures \cite{betzel2016generative,betzel2017generative}. We showed that these classic generative models failed to encode developmental-related characteristics, as none of their model parameters exhibited substantial predictive ability for brain age. Previous studies have also indicated that two-parameter generative models only explain about half of the structural connections \cite{chen2017features}, omitting crucial generative factors such as biophysics of functional integration \cite{woolrich2013biophysical}, cortical cytoarchitecture \cite{beul2015predictive}, and other regional characteristics \cite{faskowitz2018weighted,shen2020graph}. Rather than exhaustively examining all relevant generative factors,  BRIDGE tackled the complexity of connectome data by implicitly encoding brain connectivity and automatically learning the latent generative features, enabling robust and efficient characterization of neurotypical development.

We created region-wise neurodivergence maps to quantify localized neurodevelopmental divergence from neurotypical trajectories. These maps effectively highlighted the heterogeneity present at the individual level. Consistent with previous findings \cite{bethlehem2020normative, zabihi2019dissecting}, autistic children exhibited heterogeneous regional connectivity patterns. Generally, more pronounced autistic traits in ADOS were associated with increased regional neurodivergence and broader regional involvement. We observed connectivity differences in the medial visual area, lateral prefrontal cortex, frontal eye fields, and somatomotor area. These divergent regional patterns are likely related to autistic features associated with visual-motor integration \cite{lewis2017emergence, dakin2005vagaries, hubl2003functional}, executive control \cite{hughes1994evidence}, and multisensory function \cite{marco2011sensory, ben2009meta}.

Moreover, we highlighted the clinical significance of the global neurodivergence scores. Out of all models tested, global neurodivergence scores derived from the BRIDGE model correlated most strongly with clinical assessments for autism. This suggested that our BRIDGE model effectively captured neurotypical development of structural connectomes, allowing for more robust detection and quantification of neurodivergence. Previous studies on autism have typically relied on other neuroanatomical features, such as cortical thickness \cite{bethlehem2020normative, zabihi2019dissecting} and white matter integrity measures \cite{tuncc2019deviation} to quantify neurodivergence. Our study is the first to adapt normative modeling for structural connectomes. While we demonstrated the utility of normative modeling with BRIDGE for studying divergent connectome patterns in autism, our model can also serve as a reference for developing neurodivergence markers for other neurodevelopmental disorders, facilitating more personalized diagnosis and intervention in various clinical settings.

In addition to providing individualized biomarkers, normative modeling offers several advantages over group-level studies that merit consideration. It relies only neurotypical data for model training, capitalizing on the greater availability of neurotypical datasets compared to those with disorders. It also provides a way to leverage the considerable amount of connectome data that are publicly available, such as UK Biobank \cite{sudlow2015uk,miller2016multimodal} and the Human Connectome Project \cite{van2013wu}. Our findings showed that pre-training the model parameters on a larger neurotypical cohort significantly bolstered the accuracy of modeling and predicting brain age for a smaller cohort. This suggest that such an approach could provide a viable solution when training a normative model with limited sample sizes to mitigate overfitting pitfalls. Moreover, researchers can integrate datasets across multiple data sources to create a more robust reference trajectory \cite{ceriotti2007prerequisites} for neurotypical development.

This study has a few limitations. First, given the substantial differences in structural connectomes across demographics like sex, race/ethnicity, and social determinants of health \cite{ingalhalikar2014sex,rakesh2021socioeconomic}, normative models may need to be tailored to different population groups. Although developing these models would require more data, it would enhance diversity and inclusiveness in medical research, ultimately promoting health equity. Second, while this study employed pre-training techniques, the model’s training and fine-tuning was limited to the single-site data, potentially leading to site-specific biases in derived measures. Recent advances in harmonization and domain adaptation algorithms \cite{shen2024harmonization,wang2019identifying,li2020multi,fortin2017harmonization,mirzaalian2016inter,koppers2019spherical} offer promising methods to mitigate such biases and improve the generalizability of generative and normative models through data integration.

Future studies could also incorporate additional features in normative modeling, such as white matter integrity from dMRI, anatomical data from structural MRI, or genetic factors at the nodal or edgewise levels, enabling the model to integrate multiple modalities and provide more comprehensive insights into neurodevelopmental changes across a broader spectrum of brain characteristics. In addition, BRIDGE can serve as a valuable tool for data augmentation in machine learning and deep learning studies, particularly when addressing challenges such as limited sample sizes or imbalances across different ages. While data augmentation has been widely adopted to bolster the robustness and generalization of deep learning models \cite{shorten2019survey}, there are only a handful of studies that augment connectome data \cite{barile2021data,li2021brainnetgan}. The BRIDGE model, grounded in well-recognized wiring principles, provides an innovative approach to generating connectomes that closely resemble the actual ones with desired characteristics, ensuring that the augmented data retains biological relevance with diverse attributes.

In summary, our BRIDGE framework offers a novel approach to normative modeling of connectome data. By effectively capturing age-related neurotypical variability in structural connectivity, BRIDGE facilitates the quantification of individual-level developmental neurodivergence using regionally and globally interpretable connectivity markers. BRIDGE bridges critical gaps in assessing intricate transformations in connectivity underpinnings for neurodivergent individuals, paving the way for data-driven, precision medicine approaches in neurodevelopmental health.

\section{Methods}\label{sec4}

\subsection{Datasets}\label{datasets}

\bmhead{PNC cohort} The Philadelphia Neurodevelopmental Cohort (PNC) \cite{satterthwaite2014neuroimaging} included 968 neurotypical participants (425 males, 543 females) post quality assurance, aged 8-22 years (mean=15.32, SD=3.45). Diffusion MRI data were acquired on a single 3T Siemens TIM Trio whole-body scanner using a twice-refocused spin-echo (TRSE) single-shot EPI sequence. The sequence consisted of 64 gradient directions with a b-value of 1000 $s/mm^2$ and 7 scans with a b-value of 0 $s/mm^2$.

\bmhead{CAR cohort} Participants from the Center for Autism Research (CAR) cohort \cite{ghanbari2014identifying} were recruited at the Children's Hospital of Philadelphia (CHOP). A total of 229 autism children (190 males, 39 females) and 196 neurotypical participants (158 males, 38 females) were included in this study, with ages ranging from 6 to 19 years (mean=12.46, SD=2.72 for autism; mean=12.37, SD=2.96 for neurotypicals). The two diagnostic groups did not significantly differ in age. The diagnostic labels were confirmed using gold standard diagnostic instruments (ADOS \cite{lord2000autism} and ADI-R \cite{lord1994autism}), with expert consensus on clinical judgment by two independent psychologists following Collaborative Programs of Excellence in Autism (CPEA) diagnostic guidelines. Autism-related behavioral features were quantified through parent rating forms and clinical assessments by psychologists. The intelligence quotient (IQ) of each participant was measured using DAS-II \cite{elliott1990differential}. Diffusion MRI data were obtained on a Siemens 3T Verio scanner, using a 32-channel head coil and a single shot spin-echo, echo-planar sequence with $TR/TE =\num{11000}/76 ms$, b-value of 1000 $s/mm^2$, 30 gradient directions, 2 $mm$ slice thickness, 80 slices and $FOV=256\times256 mm^2$.

\subsection{Data preparation and preprocessing}
We preprocessed high-resolution T1-weighted structural images using the FreeSurfer 5.3.0 recon-all pipeline (http://surfer.nmr.mgh.harvard.edu) \cite{fischl2012freesurfer} and diffusion-weighted MRI using the MRtrix3 pipeline \cite{tournier2019mrtrix3}. The T1-weighted images were registered to the fractional anisotropy (FA) maps of each subject using a rigid registration, followed by a deformable SyN registration in ANTs \cite{avants2008symmetric} with the deformation constrained to the anterior-posterior direction to correct for the EPI distortions in the diffusion MRI.

\subsection{Construction of structural connectomes}
 We parcellated the brain into 100 regions of interest (ROIs) from the Schaefer atlas \cite{schaefer2018local}. Probabilistic fiber tracking \cite{behrens2003characterization} was performed in MRtrix3 using the iFOD2 algorithm \cite{tournier2010improved} with 250 seeds placed at random inside each voxel of the grey-matter white-matter interface (GMWMI). We uses an angle curvature threshold of 60\degree, step size of 1 $mm$, and minimum and maximum length thresholds of 25 $mm$ and 250 $mm$, respectively. We created structural connectome matrix $\bm{A} = (a_{mn})\in \mathbb{R}^{100\times100}$ using the number of streamlines intersecting pairwise combinations of brain ROIs. The connectivity strength was then normalized by dividing the number of streamlines by the total GMWMI volume of each subject. Self-connections were removed. Meanwhile, we estimated the geometric distance between each pair of brain regions by averaging the mean streamline length across all neurotypical participants in the dataset, which was used to define the cost matrix of each connection.

\subsection{Modeling neurotypical development with BRIDGE} \label{bridge}
We proposed the BRIDGE model that utilized a graph-based variational autoencoder (VAE) \cite{kipf2016variational} to characterize the neurotypical developmental trajectories of structural connectomes, as illustrated in Fig. \ref{F1}. BRIDGE took vectorized structural connectome (the upper triangular part of the matrix) as input and mapped it into lower-dimensional latent features with an encoder. The model then inferred the generation process of structural connectomes through a decoder, with the latent features capturing age-related information via a brain age predictor.

The encoder $e_\phi$ in the BRIDGE model employed a 3-layer fully connected network (FCN, with 80 and 40 hidden channels) to map the vectorized connectome matrix into a 40-dimensional latent space, estimating the mean $\bm{\mu}$ and the variance $\bm{\sigma}$ of each latent feature vector $\bm{z} \in \mathbb{R}^{40 \times 1}$. The latent features were then obtained using the reparameterization trick \cite{kingma2013auto}, where we sampled a Gaussian variable $\bm{\epsilon} \propto \mathcal{N}(0, \bm{I})$ to compute latent features $\bm{z}=\bm{\mu}+\bm{\sigma}*\bm{\epsilon}$. Brain age prediction was performed on these latent features, with a linear layer $y = {\bf z}^T {\bf \beta} + \alpha$ to parameterize the relationship between latent features $\bm{z}$ and age $y$, enabling the model to extract age-related developmental information from the vectorized structural connectome.

The conventional VAE does not impose constraints in the decoder, making it prone to overfitting, especially when characterizing high-dimensional structural connectome data with limited training samples. Additionally, the absence of constraints in modeling structural connectomes may hinder the effective capture of meaningful biological and mechanistic features, as human brain development follows specific physical and biological principles. To address this, the BRIDGE decoder adopted two well-established wiring constraints from neuroscience, the cost minimization rule and the Hebbian learning rule, to guide the generation process of structural connectomes.

Given that the human brain is physically constrained within a limited 3-dimensional volume, with long-range connections incurring high material and metabolic costs \cite{lynn2019physics, barthelemy2011spatial}, the cost minimization rule posits that the probability of forming connections between region $v_i$ and region $v_j$ decreases exponentially with the length of the connection $D_{v_i, v_j}$. We modeled this process with a 1-layer fully connected network (FCN) to estimate a learnable parameter $\eta$ from the latent features, which modulates the effects of geometric distance on connection formation. The Hebbian learning rule \cite{hebb1949organization}, as evidenced in both synaptic \cite{turrigiano2000hebb,brown2003legacy} and myelinated connections \cite{lazari2022hebbian}, consolidates connections between regions sharing similar firing configurations. To model this effect, we first mapped the latent features $\bm{z}$ to each brain region using a 1-layer FCN to create regional embeddings $\bm{z'} \in \mathbb{R}^{100 \times 1}$, followed by a 2-layer graph convolutional network (GCN) \cite{kipf2016semi} with 8 hidden channels and 16 output channels to generate regional features of brain configurations $\bm{x} \in \mathbb{R}^{100 \times 16}$. The Hebbian learning rule was formulated using the inner product $<\bm{x}_{v_i}, \bm{x}_{v_j}>$ with the regional features $\bm{x}_{v_i}$ and $\bm{x}_{v_j}$ inferred by GCNs.

The above two wiring constraints encourage the adaptation of connectivity topology to support various brain capabilities while maintaining relative low running costs \cite{petersen2015brain,park2013structural}, allowing the BRIDGE model to effectively characterize the neurotypical trajectory of structural connectomes. The overall probability $ P_{v_i,v_j}$ of forming connections between region $v_i$ and region $v_j$ was formulated as
\begin{equation}
    P_{v_i,v_j} \propto D_{v_i,v_j}^{-\eta} \sigma(<\bm{x}_{v_i}, \bm{ x}_{v_j}>)
\end{equation}
where $\sigma(\cdot)$ represents a non-linear sigmoid function for obtaining positive probability values.

The BRIDGE model was trained end-to-end by optimizing the evidence lower bound (ELBO) function for the graph VAE, which combines the encoder, biologically constrained decoder, and brain age predictor. Detailed mathematical derivations were provided in the Supplementary Materials. We used the Adam optimizer with a learning rate of $5 \times 10^{-4}$, weight decay of $10^{-4}$, 200 epochs, and batch size of 64 for training. The training process was conducted only on neurotypical participants to capture neurotypical developmental trajectories of structural connectomes within the latent space. The structural connectome of any individual can be then mapped onto the same coordinate system for assessing neurodivergence.

\subsection{Alternative models for comparison} \label{other}

\bmhead{GCN-VAE} The GCN-VAE model employed a graph-based VAE architecture to encode the developmental information in structural connectomes, with similar neural network hyper-parameters as the BRIDGE model. However, unlike BRIDGE, the GCN-VAE decoder did not impose wiring constraints when generating structural connectomes. Instead, after mapping the latent features $\bm{z}$ to the regional embeddings $\bm{z'}$, GCN-VAE simply employed a 2-layer GCN with 8 hidden channels and 100 output channels to produce a matrix $\bm{x} \in \mathbb{R}^{100\times100}$. The generated connectome was then computed as $\Tilde{\bm{A}} = \sigma(\bm{x}^T + \bm{x})$ to ensure symmetry, where $\sigma(\cdot)$ represents the sigmoid function to enforce positive edge values. The age-related developmental changes were encoded through brain age prediction based on the latent features $\bm{z}$ extracted by the GCN-VAE model.

\bmhead{FC-VAE} The FC-VAE model utilized a standard VAE architecture with neural network hyperparameters similar to those of GCN-VAE, but replacing graph convolutional layers with fully connected layers. Specifically, after mapping the vectorized structural connectome to the latent features $\bm{z} \in \mathbb{R}^{40 \times 1}$ with a 3-layer FCN (with 80 and 40 hidden channels), the FC-VAE decoder employed a symmetric 3-layer FCN (with 40 and 80 hidden channels) to reconstruct the vectorized structural connectome from the latent features and subsequently transformed this vectorized output into matrix form. The age-related information was captured by predicting the brain age based on the latent features $\bm{z}$ extracted by the FC-VAE.

\bmhead{MLP} The MLP model mapped the vectorized structural connectome to the latent features $\bm{z} \in \mathbb{R}^{40 \times 1}$ using with a 3-layer FCN (with 80 and 40 hidden channels). These latent features were then utilized to predict connectivity-based brain age through a linear layer, computed as $y = {\bf z}^T {\bf \beta} + \alpha$. Unlike three VAE-based models, MLP did not model the generation of structural connectomes when extracting age-related information.

\bmhead{LR-PCA} The LR-PCA first reduced the dimensionality of the vectorized structural connectome using principal component analysis (PCA), retaining 80\% of the data's variance. The top principal components were then utilized as latent features of structural connectomes to predict chronological age through linear regression, thereby modeling neurotypical developmental changes. No graph-based topological information of connectomes was encoded in LR-PCA.

\bmhead{SVR-PCA} The SVR-PCA encoded the vectorized structural connectome into lower-dimensional latent features using PCA with preserving 80\% of the variance. A support vector regression (SVR) was then employed to model the relationship between latent features and the participant's chronological age during neurotypical development. No graph-based topological information was encoded in SVR-PCA.

\bmhead{Classic generative models}
We assessed 13 classic generative models from prior studies \cite{betzel2016generative,betzel2017generative}, which generated structural connectomes using the two-parameter model with pre-defined wiring rules. The wiring probability for each connection was formulated using a general form:
\begin{equation}
    P_{v_i,v_j} \propto D_{v_i,v_j}^{-\eta} K_{v_i,v_j}^\gamma.
\end{equation}
where $D_{v_i,v_j}$ represents the length of the connection between node $v_i$ and $v_j$, parameterized by $\eta$ to control how connection length penalizes the probability of forming a connection; and $K_{v_i,v_j}$ denotes other graph features related to connectome generation, parameterized by $\gamma$, with the choice of the $K_{v_i,v_j}$ term listed in Table S\ref{sup-table0}. Starting with a sparse seed network consisting of 35 connections that presented across all binarized structural connectomes, new connections were added one at each iteration based on the probability $P_{v_i,v_j}$, until the generated connectome reached the same sparsity level as the actual one. The model parameters $\eta$ and $\gamma$ were optimized to minimize the dissimilarity between generated connectome $\Tilde{\bm{A}}$ and actual connectome $\bm{A}$, formulated by:
\begin{equation}
    \label{eqn:energy}
    f(\bm{A}, \Tilde{\bm{A}}) = \max{(KS_k,KS_c,KS_b,KS_e)}
\end{equation}
where each term is a Kolmogorov–Smirnov (KS) statistic that measures dissimilarity between connectomes in terms of degree (k), clustering coefficient (c), betweenness centrality (b), and edge length (e) distributions.

We performed 10,000 simulations uniformly spaced across the parameter space ($0\leq\eta\leq10$, $-10\leq\gamma\leq10$) to determine the optimal $\eta$ and $\gamma$ values for characterization of each individual's structural connectome. The $\eta$ and $\gamma$ parameters were then used as features in a linear regression model to predict chronological age, thereby capturing neurotypical developmental changes.

\subsection{Evaluation of model performance}
\bmhead{Characterization of connectome} In the VAE-based models (BRIDGE, GCN-VAE and FC-VAE), as structural connectomes were characterized using their latent features, we obtained the posterior mean of latent features for each participant and computed the associated edge-wise probability matrices $ P_{v_i,v_j}$ for forming connections through the trained decoder. We randomly generated 100 structural connectome samples based on each probability matrix while maintaining the same sparsity level as the actual connectome. Model performance was evaluated by calculating the dissimilarity score between the generated and actual connectomes (as defined in Eq. \ref{eqn:energy}), averaged across the 100 samples. For 13 classic generative models, we generated 100 structural connectomes from each pair of $\eta$ and $\gamma$ parameters, and the dissimilarity scores were computed and averaged across those samples. We repeated this analysis by thresholding structural connectomes at different sparsity levels ($\rho=10\%, 30\%, 20\%, 40\%$) and compared the model performance among BRIDGE and all other listed models (excluding MLP, LR-PCA, and SVR-PCA, as they did not model the generation of brain connectivity).

\bmhead{Characterization of age-related patterns} We predicted the brain age for each participant based on their structural connectome and employed three common metrics, mean absolute error (MAE), root mean square error (RMSE), and Pearson's correlation (R), to evaluate model performance in capturing age-related patterns. The performance of brain age prediction was assessed using 5-fold cross-validation and compared among BRIDGE and all other models listed.

\bmhead{Effects of pre-training}
Considering the relatively small sample size in the CAR cohort, we leveraged the pre-trained model parameters from the PNC cohort and fine-tuned the model for the CAR cohort to enhance model performance. We assessed the model's ability to capture age-related connectome trajectories in neurotypical development for different models with and without pre-training, allowing us to evaluate the effectiveness of the pre-training approach.

\bmhead{Ablation study}
We conducted an ablation study to evaluate the contributions of various building components in BRIDGE. Specifically, we assessed the percentage changes in MAE, RMSE and Pearson correlation R in brain age prediction upon the removal of wiring constraints (BRIDGE verse GCN-VAE), graph convolutional layers (GCN-VAE verse FC-VAE), and the VAE architecture (FC-VAE verse MLP) for generating structural connectomes.

\subsection{Detection of neurodivergence with BRIDGE}
\bmhead{Regional neurodivergence map} We proposed a regional neurodivergence map to quantify the individual's divergence from the neurotypical developmental trajectory at each brain region. For each neurotypical participant, we obtained their regional embeddings $\bm{z'} \in \mathbb{R}^{100\times1}$ using the trained BRIDGE model, which encoded region-wise connectivity patterns at each dimension. Gaussian process regression \cite{rasmussen2003gaussian} was then applied to model the age-related neurotypical trajectory at each dimension of the regional embeddings $\bm{z'}$, as well as the variability across neurotypical participants. This enabled the computation of expected neurotypical mean values $z'_{n}$ and neurotypical variance $\sigma_{n}^2$ at any given age. Following previous studies \cite{marquand2016understanding}, we computed regional neurodivergence for each neurodivergent participant using a Z score, defined as:

\begin{equation}
    Z_s = \frac{z'_{s} - z'_{n}}{\sqrt{\sigma_{s}^2 + \sigma_{n}^2}}
\end{equation}

Here, $z'_{s}$ denotes the predicted regional embedding for the participant, and the $\sigma_{s}^2$ represents the participant's prediction variance in the BRIDGE model. The individualized neurodivergence map was generated by quantifying divergence across all brain regions, enabling the identification of specific regions exhibiting the greatest neurodivergence in connectivity patterns, as illustrated in Fig.\ref{F1}C.

\bmhead{Global neurodivergence score} The disorder-related neurodivergence in brain connectivity can result in the predicted brain age $\Tilde{y}$ being different from the individual’s chronological age $y$. We used this brain age difference $\Tilde{y} - y$ as a proxy for each individual’s global neurodivergence score. Due to the sample size imbalance and modeling noises across different age, an age‐related bias often appears in brain age predictions \cite{cole2017predicting,peng2021accurate,smith2019estimation}. We applied a simple general linear model to adjust for age-related bias in predicted brain age when calculating the global divergence score, as recommended in previous studies \cite{liang2019investigating,tuncc2019deviation}.

\bmhead{Quantifying neurodivergence in autism} Using the trained BRIDGE model (with weights pre-trained on the PNC cohort), we derived regional neurodivergence maps and global neurodivergence scores for each autistic participant in the CAR cohort. To visualize individual heterogeneity in regional neurodivergence, we selected participants with similar or differing profiles in age, sex, IQ, and ADOS scores. We identified brain regions commonly associated with autism by averaging the region-wise neurodivergence maps across autistic participants. Finally, we evaluated the clinical relevance of global neurodivergence scores by examining their Spearman correlation with ADOS scores and Pearson correlation with IQ.


\bibliography{sn-bibliography}

\backmatter

\section*{Acknowledgments}
The research was supported by National Institutes of Health (NIH) grant 1-R01 MH117807-01A1. We gratefully acknowledge the efforts of all the researchers and staffs who contributed to data collection and management of the Philadelphia Neurodevelopmental Cohort (PNC) and the Center for Autism Research (CAR) cohort.

\section*{Declarations}
The authors declare no competing interests.

\section*{Supplementary information}\label{secA1}

\href{run:./sn-supplementary.pdf}{Supplementary information}

\end{document}


\title[Article Title]{Supplementary Information for Parsing altered brain connectivity in neurodevelopmental disorders by integrating graph-based normative modeling and deep generative networks}

\maketitle

\newpage
\section{Supplementary Methods}
\subsection{Derivation of evidence lower bound in VAE-based models}\label{sec1}

From a probabilistic perspective, the encoder of the VAE-based models (BRIDGE, GCN-VAE, FC-VAE) approximated the posterior distribution of latent features $q_\phi(\bm{z}|\bm{A},y)$ given observed vectorized structual connectomes and their age $y$. Then, the decoder and the brain age predictor estimated the conditional posterior distribution of the latent features, denoted as $p_\theta(\bm{z}|\bm{A}, y)$.

We used the Kullback-Leibler (KL) divergence to quantify the distance between these two posterior distributions, which was defined as
\begin{equation}
\begin{split}
D_{KL}(q_\phi(\bm{z}|\bm{A}, y)||p_\theta(\bm{z}|\bm{A}, y)) &= - E_{\bm{z} \sim q_\phi(\bm{z}|\bm{A}, y)} [\log\frac{p_\theta(\bm{z}|\bm{A}, y)}{q_\phi(\bm{z}|\bm{A}, y)}]\\
&= - E_{\bm{z} \sim q_\phi(\bm{z}|\bm{A}, y)} [\log\frac{p_\theta(\bm{z}, y|\bm{A})}{q_\phi(\bm{z}|\bm{A}, y)} - \log p_\theta(y|\bm{A})]\\
&= - E_{\bm{z} \sim q_\phi(\bm{z}|\bm{A}, y)} [\log\frac{p_\theta(\bm{z}, y|\bm{A})}{q_\phi(\bm{z}|\bm{A}, y)}] + \log p_\theta(y|\bm{A})
\end{split}
\end{equation}
As the KL divergence of two posterior distributions was non-negative, we computed the evidence lower bound (ELBO) of $\log p_\theta(y|\bm{A})$, expressed as 
\begin{equation}
\begin{split}
\mathcal{L}(\bm{A}, y;\theta,\phi) :&= E_{\bm{z} \sim q_\phi(\bm{z}|\bm{A}, y)} [\log\frac{p_\theta(\bm{z}, y|\bm{A})}{q_\phi(\bm{z}|\bm{A}, y)}]\\
&= E_{\bm{z} \sim q_\phi(\bm{z}|\bm{A}, y)} [\log p_\theta(y|\bm{A}, \bm{z}) + \log \frac{p_\theta(\bm{z}| \bm{A})}{q_\phi(\bm{z}|\bm{A}, y)}]\\
&= E_{\bm{z} \sim q_\phi(\bm{z}|\bm{A}, y)}[\log p_\theta(y|\bm{A}, \bm{z})] - D_{KL}(q_\phi(\bm{z}|\bm{A}, y) || p_\theta(\bm{z}| \bm{A}))
\end{split}
\end{equation}

The optimization was achieved by tweaking $\theta$ and $\phi$ to maximize the ELBO of $\log p_\theta(y|\bm{A})$, enabling the minimization of the KL divergence between two posteriors and a distribution that maximization $E_{\bm{z} \sim q_\phi(\bm{z}|\bm{A}, y)}[\log p_\theta(y|\bm{A}, \bm{z}))]$

\subsection{Description of 13 classic generative models}
We adopted a two-parameter model with pre-defined wiring rules. The wiring probability for each connection was formulated using a general form:
\begin{equation}
    P_{v_i,v_j} \propto D_{v_i,v_j}^{-\eta} K_{v_i,v_j}^\gamma.
\end{equation}
We computed $D_{v_i,v_j}$ using the average streamline length between node $v_i$ and $v_j$, which was parameterized by $\eta$ to control the penalty on connection length. The topological term $K_{v_i,v_j}$ was pre-defined using graph features as listed in Table S\ref{table0}, parameterized by $\gamma$.

\begin{table*}[h!]
\caption{13 classic generative models with pre-defined topological terms}
\begin{adjustbox}{width=0.7\textwidth,center}
    \begin{tabular}{c|c|l}
        Model & $K_{v_i,v_j}$  & Category \\
        \toprule
        geometric & 1 & \multirow{1}{0.25\linewidth}{Use constant value}\\
        \hline
        deg-diff & $|k_{v_i} - k_{v_j}|$ & \multirow{5}{0.25\linewidth}{Based on node degree $k_{v_i}$ and $k_{v_j}$}\\
        \hhline{--~}
        deg-min & $\min{(k_{v_i}, k_{v_j})}$ &  \\
        \hhline{--~}
        deg-prod & $k_{v_i}k_{v_j}$ &  \\
        \hhline{--~}
        deg-max & $\max{(k_{v_i}, k_{v_j})}$ &  \\
        \hhline{--~}
        deg-avg & $(k_{v_i} + k_{v_j})/2$ &  \\
        \hline
        clu-diff & $|c_{v_i} - c_{v_j}|$ & \multirow{5}{0.25\linewidth}{Based on node clustering coefficient $c_{v_i}$ and $c_{v_j}$}\\
        \hhline{--~}
        clu-min & $\min{(c_{v_i}, c_{v_j})}$ &  \\
        \hhline{--~}
        clu-prod & $c_{v_i}c_{v_j}$ &  \\
        \hhline{--~}
        clu-max & $\max{(c_{v_i}, c_{v_j})}$ &  \\
        \hhline{--~}
        clu-avg & $(c_{v_i} + c_{v_j})/2$ &  \\
        \hline
        neighbor & $|\Gamma_{v_i\backslash v_j} \cap \Gamma_{v_j\backslash v_i}|$ & \multirow{2}{0.25\linewidth}{Based on common neighbors$^1$}\\
        \hhline{--~}
        matching & $\frac{|\Gamma_{v_i\backslash v_j} \cap \Gamma_{v_j\backslash v_i}|}{|\Gamma_{v_i\backslash v_j} \cup \Gamma_{v_j\backslash v_i}|}$ &  \\
        \bottomrule
    \end{tabular}
    \label{table0}
\end{adjustbox}
\footnotesize{$^1$ $\Gamma_v = \{u: a_{uv} = 1\}$ represents a set of neighboring nodes for node $v$, and $\Gamma_{v\backslash u}$ denotes neighboring nodes of node $v$ but excluding node $u$.}
\end{table*}

\clearpage
\section{Supplementary Results}
\subsection{Evaluation of classic generative models for neurotypicals}
\begin{table*}[h!]
\caption{Correlation analysis on neurotypicals between chronological age and features from classic generative models}
\begin{adjustbox}{width=0.7\textwidth,center}
    \begin{tabular}{c|c|c|c|c}
        \multicolumn{1}{c|}{} & \multicolumn{2}{c|}{PNC} & \multicolumn{2}{c}{CAR}\\
        \toprule
        \multicolumn{1}{c|}{Models} & \multicolumn{1}{c|}{$\eta$} & \multicolumn{1}{c|}{$\gamma$} & \multicolumn{1}{c|}{$\eta$} & \multicolumn{1}{c|}{$\gamma$} \\
        \hline
        geometric & 0.0015 & --- & -0.0041 & --- \\
        \hline
        deg-diff & 0.0171 & -0.0893 & -0.0781 & 0.0061 \\
        \hline
        deg-min & -0.0034 & -0.1674* & -0.1722 & -0.2132 \\
        \hline
        deg-prod & -0.1112* & -0.1824* & -0.1357 & -0.1774 \\
        \hline
        deg-max & -0.1226* & -0.1359* & -0.1996 & -0.0803 \\
        \hline
        deg-avg & -0.1450* & -0.1400* & -0.2437* & -0.1435 \\
        \hline
        clu-diff & 0.1596* & 0.1013* & 0.1581 & 0.1624 \\
        \hline
        clu-min & -0.0703 & 0.0293 & -0.1112 & 0.0562 \\
        \hline
        clu-prod & -0.0209 & 0.0136 & -0.0342 & -0.0216 \\
        \hline
        clu-max & 0.0397 & 0.1617* & 0.0370 & 0.1835 \\
        \hline
        clu-avg & 0.0224 & 0.1212* & 0.2086 & 0.0573 \\
        \hline
        neighbor & 0.0206 & -0.0552 & 0.1268 & -0.0885 \\
        \hline
        matching & 0.06472 & 0.0067 & 0.0183 & -0.0418 \\
        \bottomrule
    \end{tabular}
    \label{table1}
\end{adjustbox}
\footnotesize{$^1$Significant correlations (FWER corrected using the Holm-Šídák method) are indicated with asterisks.}
\end{table*}

\begin{table*}[h!]
    \caption{Brain age prediction on neurotypicals with features from classic generative models}
    \centering 
    \begin{tabular}{c|c|c|c|c|c|c}
        \multicolumn{1}{c|}{} & \multicolumn{3}{c|}{PNC} & \multicolumn{3}{c}{CAR}\\
        \toprule
        \multicolumn{1}{c|}{Models} & \multicolumn{1}{c|}{MAE} & \multicolumn{1}{c|}{RMSE} & \multicolumn{1}{c|}{R} & \multicolumn{1}{c|}{MAE} & \multicolumn{1}{c|}{RMSE} & \multicolumn{1}{c}{R} \\
        \hline
        geometric & 2.9034 & 3.4444 & -0.0677 & 2.5075 & 2.9779 & -0.0612 \\
        \hline
        deg-diff & 2.8904 & 3.4321 & 0.0706 & 2.5224 & 2.9870 & 0.0591 \\
        \hline
        deg-min & 2.8505 & 3.3846 & 0.1832 & 2.5006 & 2.9665 & 0.1759 \\
        \hline
        deg-prod & 2.8531 & 3.3853 & 0.1773 & 2.479 & 2.9636 & 0.1619 \\
        \hline
        deg-max & 2.8609 & 3.4073 & 0.1378 & 2.485 & 2.9427 & 0.1384 \\
        \hline
        deg-avg & 2.8538 & 3.4014 & 0.1503 & 2.4471 & 2.9094 & 0.2151 \\
        \hline
        clu-diff & 2.8596 & 3.4025 & 0.152 & 2.5005 & 2.9422 & 0.1434 \\
        \hline
        clu-min & 2.8949 & 3.4389 & 0.0423 & 2.4924 & 2.9661 & 0.0405 \\
        \hline
        clu-prod & 2.9084 & 3.4508 & -0.0754 & 2.5101 & 2.9859 & -0.1085 \\
        \hline
        clu-max & 2.8362 & 3.3704 & 0.1997 & 2.5094 & 2.9439 & 0.1717 \\
        \hline
        clu-avg & 2.8694 & 3.3932 & 0.166 & 2.4622 & 2.9222 & 0.2070 \\
        \hline
        neighbor & 2.895 & 3.4352 & 0.0558 & 2.5184 & 2.9725 & 0.0623 \\
        \hline
        matching & 2.8992 & 3.4355 & 0.0579 & 2.4906 & 2.954 & 0.0563 \\
        \bottomrule
    \end{tabular}
    \label{table2}
\end{table*}

\clearpage
\subsection{Evaluation on autism}
\begin{table*}[h]
\caption{Brain age prediction on autism}
\begin{adjustbox}{width=0.6\textwidth,center}
    \begin{tabular}{c|c|c|c}
        \multicolumn{1}{c|}{Models} & \multicolumn{1}{c|}{MAE} & \multicolumn{1}{c|}{RMSE} & \multicolumn{1}{c}{R}\\
        \toprule
        $\bar{y}_{age}$ (Baseline) & 2.2416 & 2.7209 & --- \\
        \hline
        LR-PCA & 1.7928 & 2.3426 & 0.6125 \\
        \hline
        SVR-PCA & 1.7570 & 2.2484 & 0.6460 \\
        \hline
        MLP & 1.5981 & 2.0357 & 0.6866 \\
        \hline
        FC-VAE & 1.5708 & 2.0069 & 0.6953 \\
        \hline
        GCN-VAE & 1.5629 & 2.0113 & 0.6992 \\
        \hline
        BRIDGE & \textbf{1.5483} & \textbf{1.9839} & \textbf{0.7031} \\
        \bottomrule
    \end{tabular}
    \label{table3}
\end{adjustbox}
\footnotesize{$^1$The prediction model were trained on structural connectivity of neurotypical participants in the CAR dataset (w/ pretrain on the PNC dataset).}
\end{table*}